\documentclass[journal]{IEEEtran}
\usepackage{amsmath,amsfonts,amsthm}
\usepackage{array}
\usepackage[caption=false,font=normalsize,labelfont=sf,textfont=sf]{subfig}
\usepackage{textcomp}
\usepackage{stfloats}
\usepackage{url}
\usepackage{verbatim}
\usepackage{graphicx}
\usepackage{xcolor}
\usepackage{nomencl}
\usepackage{etoolbox}
\usepackage{ifthen}
\usepackage{booktabs}
\usepackage[ruled]{algorithm2e}

\usepackage{multirow}
\usepackage{bm}
\usepackage{amsmath}
\usepackage{makecell}
\usepackage{enumitem}

\allowdisplaybreaks 
\usepackage[subtle,tracking=normal]{savetrees}

\usepackage{cite}

\def\BibTeX{{\rm B\kern-.05em{\sc i\kern-.025em b}\kern-.08em
    T\kern-.1667em\lower.7ex\hbox{E}\kern-.125emX}}
\usepackage{balance}

\begin{document}

\bstctlcite{IEEEexample:BSTcontrol}
\title{$\text{LiFePO}_{\text{4}}$ Battery SOC Estimation under \\ OCV-SOC Curve Error Based on\\ Adaptive Multi-Model Kalman Filter}

\author{\IEEEauthorblockN{Daniyaer Paizulamu, \textit{Student Member, IEEE}}, \IEEEauthorblockN{Lin Cheng, \textit{Senior Member, IEEE}},\\ \IEEEauthorblockN{Yingrui Zhuang, \textit{Student Member, IEEE}}, \IEEEauthorblockN{Helin Xu}, \IEEEauthorblockN{Ning Qi}, \textit{Member, IEEE}, and \IEEEauthorblockN{Song Ci}, \textit{Fellow, IEEE}


\thanks{
This work was supported by the National Key Research and Development Program of China (No. 2023YFB2407900), 
National Natural Science Foundation of China (No. 52037006).
(\textit{Corresponding author: Daniyaer Paizulamu.})

Daniyaer Paizulamu, Lin Cheng, Yingrui Zhuang, Helin Xu and Song Ci are with the State Key Laboratory of Power System Operation and Control, Department of Electrical Engineering, Tsinghua University, Beijing 100084, China (e-mail: dnyepzlm22@mails.tsinghua.edu.cn; chenglin@mail.tsinghua.edu.cn; zyr21@mails.tsinghua.edu.cn; xhl20@mails.\\tsinghua.edu.cn; sci@mail.tsinghua.edu.cn).

Ning Qi is with the Department of Earth and Environmental Engineering, Columbia University, New York, NY 10027, USA (e-mail: nq2176@columbia.edu). 
}
\vspace{-1.5em}
}

\markboth{IEEE TRANSACTIONS ON TRANSPORTATION ELECTRIFICATION,~Vol.~X, No.~X, XX Month~2024}
{How to Use the IEEEtran \LaTeX \ Templates}

\maketitle

\IEEEaftertitletext{\vspace{-1\baselineskip}}

\begin{abstract}
$\text{LiFePO}_{\text{4}}$ batteries are widely used in electric vehicles and energy storage systems due to long cycle life and high safety performance. However, the OCV-SOC curve (OSC) of these batteries features a long plateau region, making state of charge (SOC) estimation highly sensitive to OSC error, which arises due to aging and temperature. To address this, we propose an SOC estimation method that accounts for error in OSC.
First, we establish battery equivalent circuit model (ECM) and introduce a parameters identification algorithm based on adaptive recursive least squares. Next, we derive the relationship between the innovation's cross-correlation matrix (CCM)/ auto-correlation matrix (ACM) of the Kalman filter and the OSC error. We then develop an adaptive multi-model Kalman filter (AMMKF), which dynamically adjusts the measurement model parameters of each filter based on the sign of the OSC error. By assigning a probability to each filter according to its predicted voltage distribution function, the optimal filter is selected.
The proposed method is tested under various OSC error types and  operating conditions. Results demonstrate that the proposed method achieves high accuracy and robustness, with an RMSE of less than 3\%, which is more than 10\% lower than the estimation error of traditional method.
\end{abstract}
\mbox{}

\section{Introduction}\label{Introduction}
\IEEEPARstart{L}{ithium-ion} batteries are extensively used in electric vehicles and energy storage systems~\cite{LFP}. Among these, $\mathrm{LiFePO_4}$ batteries are gaining market share due to their high safety, long cycle life, and low cost. The accuracy of battery SOC estimation directly impacts the evaluation of battery power and remaining energy~\cite{joint}. Moreover, precise SOC estimation is crucial for preventing overcharging and overdischarging, making it essential for maintaining battery consistency and ensuring safety~\cite{balance}.

Among the SOC estimation methods for lithium-ion batteries, the Coulomb Counting method~\cite{AH} is the easiest to implement in Battery Management Systems (BMS). However, it cannot correct initial estimation errors and tends to accumulate errors over time. The Open Circuit Voltage (OCV) method~\cite{O-C-V} estimates SOC by referencing an OCV-SOC table, but it requires the battery to remain stationary for an extended period, making it unsuitable for online estimation. Data-driven methods~\cite{datadriven} require a large dataset under various operating conditions for specific batteries. Recent research on SOC estimation has primarily focused on model-based methods and hybrid approaches that combine data-driven techniques with models~\cite{fused1},~\cite{fused2}. Model-based estimation methods primarily rely on filters, including Extended Kalman Filters (EKF)\cite{EKF}, Unscented Kalman Filters (UKF)\cite{UKF}, Dual Extended Kalman Filters (DEKF)\cite{DEKF}, H-infinity filters\cite{H-infinity}, and particle filters~\cite{particle}. 

OSC error has a significant impact on SOC estimation results~\cite{Curve-effect}. The OSC is primarily influenced by battery aging and temperature fluctuations~\cite{wang2019state}, with a 1 mV OSC error potentially leading to nearly a 5\% SOC error during the plateau period~\cite{5mv}. One common approach to addressing OSC errors is to obtain OSC data at various aging levels and temperatures~\cite{OCV-T}. Another approach involves developing methods to quickly identify the OSC under the current battery state. A direct and rapid OSC capture method, based on a one-cycle bipolar-current pulse, is proposed in~\cite{fastOCV1}. By analyzing the battery relaxation model, it is shown that this pulse excitation effectively shortens the relaxation time needed for OCV measurement. The trend of OSC changes can also be predicted. A prediction method that incorporates a correction mechanism is proposed in~\cite{OCVpredict}, where a feedback system corrects subsequent curves based on the prediction errors of previous curves. The electrochemical mechanisms behind OSC curve deformation are explored in~\cite{OCVLSTM}, which proposes an OCV estimation method based on partial charging data, utilizing an LSTM model with a many-to-one structure.

To improve the SOC estimation accuracy of $\mathrm{LiFePO_4}$ batteries, model-based research has focused on several key areas: enhancing battery modeling~\cite{10188385},\cite{model2}, improving the accuracy of ECM parameter estimation\cite{8924911}, refining filter-based algorithms, and modifying measurement models. A novel Information Appraisal Procedure based on Recursive Least Squares (IAP-RLS) is introduced in~\cite{du2021information}, which mitigates the impact of noise on parameter estimation. Unlike traditional RLS, IAP-RLS does not require continuous excitation. Moldel-based algorithms aim to enhance filter performance, particularly by improving noise resistance and reducing the impact of parameter estimation errors. An adaptive method for updating EKF parameters is proposed in~\cite{xiong2023state}, which effectively prevents SOC estimation divergence during the plateau period compared to traditional filters. To address the hysteresis phenomenon and noise interference caused by inaccurate voltage and current measurements, an Invariant-Imbedding-Method (IIM) was applied in SOC estimation in~\cite{DONG2016163}. This approach mitigates the effects of hysteresis and eliminates errors introduced by sensor noise.

In addition to innovating estimator algorithms, adjusting the OCV-SOC curve is also an effective approach to improve estimation accuracy during the plateau period. In~\cite{zhang2023novel}, a pseudo-OCV modeling method is proposed, which optimizes estimation by balancing the increase in the OSC slope during the plateau period against voltage error. The relationship between SOC and charging voltage is analyzed in~\cite{wang2019state}, where the OSC in the measurement equation of the DEKF is replaced with charging voltage, resulting in a new estimator that demonstrates strong robustness against noise in the plateau period. An Incremental Capacity (IC) curve-based SOC estimation method is proposed in~\cite{IC}, which can correct the cumulative error of the Ampere-hour integration method in high-rate charging scenarios. The EKF and Particle Filter methods based on the relationship between differential voltage and SOC are developed in~\cite{DV}. However, these methods still depend on relatively accurate external characteristic curves. When there is an OSC error during the plateau region, the state estimation results are likely to converge to a point with a significant deviation.

The aforementioned model-based SOC estimation algorithms rely heavily on accurate OSC, whether derived from multiple experiments or fast estimation methods. However, this approach imposes a substantial workload on battery systems containing numerous cells, making it necessary to explore SOC estimation methods that can function effectively with OSC errors. In this paper, we establish an AMMKF based on the analysis of innovation's CCM and ACM. This approach enables accurate SOC estimation for $\mathrm{LiFePO_4}$ batteries even with OSC error, thereby reducing the Battery Management System's (BMS) dependence on real-time OSC curves. Specifically, our contributions are as follows:

\begin{enumerate}
    \item \textbf{\textit{ECM Parameters Estimation Based on ARLS:}} 
        We introduced an adaptive forgetting factor in the  RLS method, which is dynamically adjusted based on feedback from SOC estimation results. This enables the method to sensitively capture parameters change, particularly as the battery approaches full charge or a low state of charge.
    \item \textbf{\textit{Innovation CCM and ACM Analysis:}} 
        We discovered that when the KF measurement equation incorporates an OSC with error, the analysis of the innovation's CCM and ACM can effectively characterize the deviation between this original OSC and the actual OSC of the battery under different aging states and operating temperatures.
    \item \textbf{\textit{SOC Estimation Based on AMMKF:}}
        We developed an Adaptive Multi-Model Kalman Filter that dynamically adjusts the measurement model parameters of the filters based on the analysis of innovation CCM and ACM results. The probability of each filter's validity is evaluated based on the distribution characteristics of their predicted voltage, and the optimal filter is selected.
\end{enumerate}

The remainder of the paper is organized as follows.
Section~\ref{Battery} establishes the battery model and proposes the ARLS parameters identification algorithm. Section~\ref{Method} derives the characterization relationship between the innovation CCM, ACM, and OSC error, leading to the development of the AMMKF-based SOC estimation method. Section~\ref{Results} presents the performance test results of the proposed method on a public experimental dataset. Finally, the conclusions are summarized in Section~\ref{Conclusion}.

\section{Battery Modeling and Online Parameters Identification Method}\label{Battery}

\subsection{Battery Model}
\begin{figure}[t] 
    \vspace{-0.2cm}
    \centerline{\includegraphics[width=0.95\columnwidth]{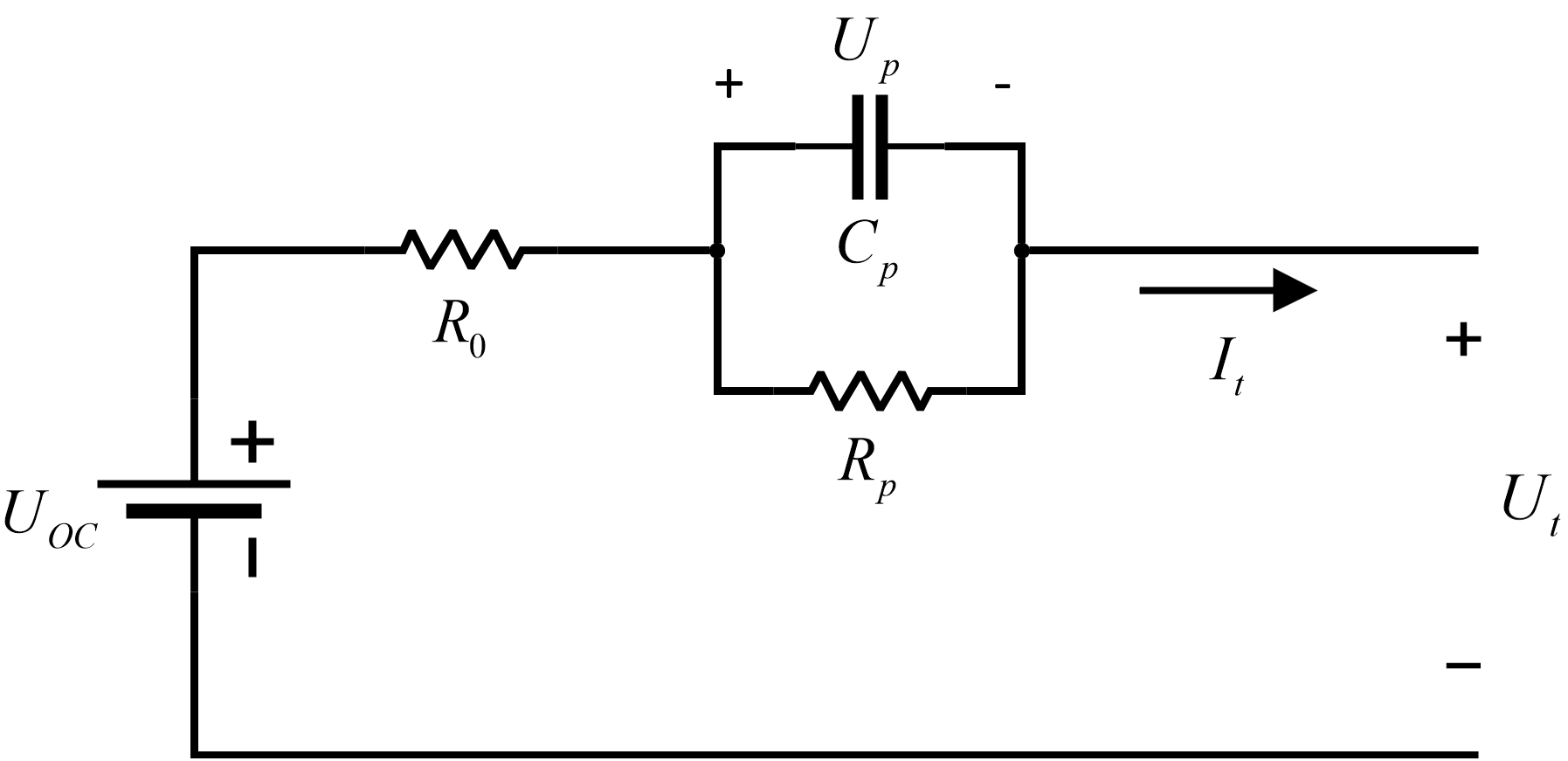}}
    \caption{Circuit diagram of the Thevenin model}
    \label{ECM}
    \vspace{-0.5cm}
\end{figure}

The ECM of a battery can effectively describe both the static and dynamic characteristics of a lithium-ion battery. From an electrochemical perspective, the resistance and capacitance elements in the model accurately reflect the influence of the electrochemical reaction process, electrolyte distribution, and polarization effects on the battery's terminal voltage. The accuracy of the model is determined by the number of RC links included in the ECM.
Considering the balance between model accuracy and complexity, this paper selects the first-order Thevenin model for analysis, as shown in Fig.~\ref{ECM}. \(R_{0}\) represents the ohmic internal resistance of the battery, which accounts for the sudden change in terminal voltage \(U_t\) when switching between charging and discharging states. The elements \(R_{p}\) and \(C_{p}\) describe the battery's polarization effect, capturing the gradual change in \(U_t\) over time. The open-circuit voltage \(U_{oc}\) is correlated with the battery's SOC under any operating condition, reflecting the battery's static characteristics. The polarization voltage across the RC link is denoted by \(U_p\). The variable \(I_L\) represents the battery's charging or discharging current, with the charging current direction considered positive.

\subsection{ARLS Parameters Identification Method}
The continuous-time state equation of the battery Thevenin equivalent circuit model is expressed as
\begin{subequations}\label{equation 1}
    \begin{align}
    \label{equation 1a}
     \frac{{d{U_p}}}{{dt}} =  - \frac{{{U_p}}}{\tau } + \frac{1}{{{C_p}}}{I_L}\\
        \label{equation 1b}
{U_p} ={U_{oc}}-U_t- {R_{\rm{0}}}{I_L}
    \end{align}
\end{subequations}

where $\tau  = {R_p}{C_p}$. The discrete form of (1) can be expressed as
\begin{subequations}\label{equation 2}
    \begin{align}
    \label{equation 2a}
{U_p}(k) = {e^{ - \frac{{\Delta t}}{\tau }}}{U_p}(k - 1) + (1 - {e^{ - \frac{{\Delta t}}{\tau }}}){R_p}{I_L}(k - 1)\\
\label{equation 2b}
{U_p}(k) = {U_{oc}}(k)-{U_t}(k) - {R_0}{I_L}(k)
    \end{align}
\end{subequations}

\noindent where \({\Delta t}\) is the sampling frequency, which is set to 1s in this paper. Substituting (2b) into (2a), \(U_{t}\) can be expressed as
\begin{equation}\label{3}
    \begin{split}
        {U_t}(k) = {U_{oc}}(k) &- {\theta _1}({U_{oc}}(k - 1) - {U_t}(k - 1)) \\
            &+ {\theta _2}{I_L}(k) + {\theta _3}{I_L}(k - 1)
    \end{split}
\end{equation}

\noindent where \({\boldsymbol{\theta} ^{\rm T}}=[{\theta _1}, {\theta _2}, {\theta _3}]=[{e^{ - \frac{{\Delta t}}{\tau }}}, -{R_0}, {e^{ - \frac{{\Delta t}}{\tau }}}{R_0} - (1 - {e^{ - \frac{{\Delta t}}{\tau }}}){R_p}]\). Similarly, we can get the expression of \({U_t}(k-1)\), and subtract it from (3) to get:
\begin{equation}\label{4}
    \begin{split}
\Delta {U_t}(k) = {\theta _1}\Delta {U_t}(k - 1) + {\theta _2}\Delta {I_L}(k) + {\theta _3}\Delta {I_L}(k - 1)\\
 + \Delta {U_{oc}}(k) - {\theta _1}\Delta {U_{oc}}(k - 1)
    \end{split}
\end{equation}

\noindent where $\Delta {I_L}(k), \Delta {U_{oc}}(k)$ represent first-order backward difference. The dynamic change of the battery open circuit voltage \(U_{oc}\) is mainly affected by battery aging and current operating conditions, so it can be described as a function of \(SOC(t)\), temperature \(T(t)\) , current capacity \(C_{n}\).Therefore, the rate of change of \(U_{oc}\) at any time is
\begin{equation}\label{Equation 5}
\frac{{d{U_{oc}}}}{{dt}} = \frac{{\partial {U_{oc}}}}{{\partial SOC}}\frac{{\partial SOC}}{{\partial t}} + \frac{{\partial {U_{oc}}}}{{\partial T}}\frac{{\partial T}}{{\partial t}}{\rm{ + }}\frac{{\partial {U_{oc}}}}{{\partial {C_n}}}\frac{{\partial {C_n}}}{{\partial t}}
\end{equation}

For the plateau region SOC $\in$ [20\%, 80\%], the value of \(\frac{{\partial {U_{oc}}}}{{\partial SOC}}\) is small. If an efficient algorithm is used to effectively shorten the convergence time of the parameter estimation results, the change in SOC can be ignored, \(\frac{{\partial SOC}}{{\partial t}} \approx {\rm{0}}\). The battery temperature changes at a low rate when in non-fault operation, so \(\frac{{\partial T}}{{\partial t}} \approx {\rm{0}}\), and the battery aging state does not change within a single cycle, so \(\frac{{\partial {C_n}}}{{\partial t}} \approx {\rm{0}}\). \(\Delta {U_{oc}}(k) - {\theta _1}\Delta {U_{oc}}(k - 1) \approx \frac{{d{U_{oc}}}}{{dt}}\Delta t \approx {\rm{0}}\). Therefore, (4) can be rewritten as
\begin{equation}\label{Equation 6}
y_k = \boldsymbol{A}_k\boldsymbol{\theta} + v_k
\end{equation}

\noindent where \(\boldsymbol{A}_k = (\Delta {U_t}(k - 1),\Delta {I_L}(k),\Delta {I_L}(k - 1))\), \({y_k} = \Delta {U_t}(k)\), $v \sim N(0,{\sigma ^2})$ is the measurement noise, and the noise at each moment is independent of each other.
\begin{algorithm}[htbp]\label{algorithm1}
    \caption{ARLS for Parameters Identification}
    \SetAlgoLined
    \SetEndCharOfAlgoLine{}
    \KwIn{The change of \(U_t\) and \(I_L\) at adjacent points.}
    \KwOut{Circuit model parameters: \(R_0\), \(R_p\), \(C_p\).}
    \SetKwBlock{StepOne}{Step 1 - Initialization}{}
    \SetKwBlock{StepTwo}{Step 2 - Update Model Gain and Error Covariance}{}
    \SetKwBlock{StepThree}{Step 3 - Update Parameters Values}{}
    \SetKw{Parallel}{parallel}
    \StepOne{
    Set initial parameters value \(\hat{\boldsymbol{\theta}}_0= \mathbb{E}(\boldsymbol{\theta})\)\\
    Set error covariance \({\boldsymbol{P}_0} = \mathbb{E}[(\boldsymbol{\theta}  - \hat{\boldsymbol{\theta}}_0){(\boldsymbol{\theta}  - \hat{\boldsymbol{\theta}}_0)^{\rm T}}]\)
    }
    \StepTwo{
     FF $\lambda_k=1-a\left|\hat{SOC}_{k-1}^+ -\frac{1}{2}\right|\frac{\hat{SOC}_{k-1}^+}{\hat{SOC}_{k-2}^+}$\\
     Gain \(\boldsymbol{K}_k = {\boldsymbol{P}_{k - 1}}\boldsymbol{A}_k^{\rm T}{({\lambda _k} + {\boldsymbol{A}_k}{\boldsymbol{P}_{k - 1}}\boldsymbol{A}_k^{\rm T})^{ - 1}}\)\\
     Error Covariance \({\boldsymbol{P}_k} = \frac{1}{{{\lambda _k}}}(\boldsymbol{I} - {\boldsymbol{K}_k}{\boldsymbol{A}_k}){\boldsymbol{P}_{k - 1}}\)
    }
    \StepThree{
    Update \(\boldsymbol{\theta}\) \(\boldsymbol{\theta}_k = \boldsymbol{\theta}_{k - 1} + \boldsymbol{K}_k (y_k - \boldsymbol{A}_k \boldsymbol{\theta}_{k - 1})\)\\
    Update Circuit Parameters
    \[[{R_0},{R_p},{C_p}] = [-{\theta _2},\frac{{{\theta _1}{\theta _2} + {\theta _3}}}{{{\theta _1}-1}},\frac{{(1-{\theta _1})\Delta t}}{{\ln {\theta _1} \times ({\theta _1}{\theta _2} + {\theta _3})}}]\]
    }
    
    \end{algorithm}
The specific algorithm flow of ARLS is shown in \textbf{Algorithm~1}. The forgetting factor \(\lambda \) is used to prevent data saturation and make the parameters estimation results more biased toward recent data and closer to the current parameters values. In the low SOC range (SOC\textless 10\%), the circuit parameters change more dramatically, so a smaller forgetting factor can be selected. After obtaining a stable value $\boldsymbol{\theta}$ through ARLS, the ECM parameters can be indirectly calculated.

\section{AMMKF-based SOC estimation method under OSC error}\label{Method}

\subsection{Single Extended Kalman Filter Model}
The Kalman filter used for SOC estimation needs to comprehensively consider the SOC change equation based on the Coulomb counting method (7) and the open circuit voltage method. Its essence is to multiply the probability distribution of the state quantity estimated by the two methods to obtain the overlapping part of the probability distribution.

\begin{equation}\label{Equation 7}
SOC(t) = SOC({t_0}) + \frac{{\eta \int_{{t_0}}^t {I(\tau )d\tau } }}{{{C_n}}}
\end{equation}

\noindent where \(\eta\) is the coulombic efficiency of the battery, which is usually set to 1, this will bring certain model errors for retired batteries. Combining (1b) with the discrete form of (7) can obtain a single discrete Kalman filter model expressed as (8), the measurement equation (open circuit voltage equation) is used to correct the SOC to avoid a large cumulative error during long-term operation.
\begin{subequations}\label{equation 8}
    \begin{align}
    \label{equation 8a}
     {\boldsymbol{x}_k} &= {\boldsymbol{F}_{k - 1}}{\boldsymbol{x}_{k - 1}} + {\boldsymbol{G}_{k - 1}}{u_{k - 1}} + {\boldsymbol{w}_{k - 1}}\\
        \label{equation 8b}
{y_k} &= {h_k}({\boldsymbol{x}_k}) + {D_k}{u_k} + {v_k}
    \end{align}
\end{subequations}
\[\boldsymbol{F}_{k - 1} = \begin{bmatrix}
1 & 0 \\
0 & \exp\left( - \frac{\Delta t}{R_{p,k-1} C_{p,k-1}} \right)
\end{bmatrix}\]
\[\boldsymbol{G}_{k - 1} = \begin{bmatrix}
 - \frac{{\Delta t}}{{{C_n}}}\\
{R_{p,k - 1}}(1 - \exp ( - \frac{{\Delta t}}{{{R_{p,k - 1}}{C_{p,k - 1}}}}))
\end{bmatrix}\]

\noindent where $\boldsymbol{x}_k = {[SOC_k{\rm{ }},{U_{p,k}}]^{\rm T}}$, ${u_k} = {I_k}$, ${h_k}({x_k}) = {U_{oc,k}} - {U_{p,k}}{\rm{  }}$,  ${D_k} =  - {R_{0,k}}$, ${\boldsymbol{w}_k}$ and ${v_k}$ represent the process noise and measurement noise, This paper assumes that they are zero-mean uncorrelated Gaussian white noise. $\mathbb{E}({\boldsymbol{w}_k}\boldsymbol{w}_j^{\rm T}) = {\boldsymbol{Q}_k}{\delta _{k - j}}$, $\mathbb{E}({v_k}v_j^{\rm T}) = {R_k}{\delta _{k - j}}$, $\mathbb{E}({w_k}v_j^{\rm T}) = 0$, ${\delta _{k - j}} = 1$ when $k = j$, otherwise ${\delta _{k - j}} = 0$. Considering the main affecting factors, OCV is expressed as ${\left. {{U_{oc,k}} = {f_k}(SO{C_k})} \right|_{{T_k},{C_n} }}$, which is a nonlinear expression. In the plateau region of the OSC, the nonlinearity is minimal, making the performance of the EKF comparable to that of the UKF. The AMMKF introduced in this paper necessitates the construction of a parameter set based on the first-order derivative of the curve. Given these considerations, the Extended Kalman Filter is selected for further analysis. The algorithm is detailed in \textbf{Algorithm~2}.

The SOC estimation method that relies on a single EKF requires a precise and known expression for the function $U_{oc}$. Using an original OSC without considering this limitation can result in significant model error, leading to inaccurate or even divergent filter estimation outcomes. These issues will be explored in the next subsection.

\begin{algorithm}[htbp]\label{algorithm2}
    \caption{EKF Approach for SOC Estimation}
    \SetAlgoLined
    \SetEndCharOfAlgoLine{}
    \KwIn{Measurement data of \(U_t\) and \(I_L\), $\boldsymbol{Q}$, $R$}
    \KwOut{Estimated value of SOC}
    \SetKwBlock{StepOne}{Step 1 - Initialization and Linearization}{}
    \SetKwBlock{StepTwo}{Step 2 - Prior Estimation}{}
    \SetKwBlock{StepThree}{Step 3 - Posterior Estimation}{}
    \StepOne{
        Step 1.1: Initialize state \(\hat{\boldsymbol{x}}_{0}^{+}= \mathbb{E}(\boldsymbol{x}_0)\)\\
        Step 1.2: Initialize the error covariance \\
        \(\boldsymbol{P}_0^ +  = \mathbb{E}[({\boldsymbol{x}_0} -\hat{\boldsymbol{x}}_{0}^{+}){({\boldsymbol{x}_0} -\hat{\boldsymbol{x}}_{0}^{+})^{\rm T}}]\)\\
        Step 1.3: Linearization of the measurement equation\\
        \(
        {y_k} = {h_k}(\hat{\boldsymbol{x}}_k^{-}) + {\left. {\frac{{\partial {h_k}}}{{\partial \boldsymbol{x}}}} \right|_{\hat{\boldsymbol{x}}_k^{-}}}(\boldsymbol{x}_k -\hat{\boldsymbol{x}}_k^{-}) + {\left. {\frac{{\partial {y_k}}}{{\partial u}}} \right|_{\hat{\boldsymbol{x}}_k^{-}}}{u_k} + {\left. {\frac{{\partial {y_k}}}{{\partial v}}} \right|_{\hat{\boldsymbol{x}}_k^{-}}}{v_k} = \boldsymbol{H}_k\boldsymbol{x}_k + {D_k}{u_k} + {v_k} + {z_k}
        \)\\
        \(
        \boldsymbol{H}_k = {\left. {\frac{{\partial {h_k}}}{{\partial \boldsymbol{x}}}} \right|_{\hat{\boldsymbol{x}}_k^{-}}} = [{\left. {\frac{{\partial {f_k}(SO{C_k})}}{{\partial SOC}}} \right|_{\hat{SOC}_k^{-}}}, - 1]
        \)\\
        \(
        {z_k} = {h_k}(\hat{\boldsymbol{x}}_k^{-}) - \boldsymbol{H}_k\hat{\boldsymbol{x}}_k^{-}
        \)
    }
    \StepTwo{
        Step 2.1 Prior state update\\
        \(\hat{\boldsymbol{x}}_k^{-}= {\boldsymbol{F}_{k - 1}}\hat{\boldsymbol{x}}_{k-1}^{+} + {\boldsymbol{G}_{k - 1}}{u_{k - 1}}\)\\
        Step 2.2 Prior error covariance update\\
        \(\boldsymbol{P}_k^ -  = {\boldsymbol{F}_{k - 1}}\boldsymbol{P}_{k - 1}^ + \boldsymbol{F}_{k - 1}^{\rm T} + {\boldsymbol{Q}_{k - 1}}\)
    }
    \StepThree{
        Step 3.1 Kalman gain\\
        \(
        {\boldsymbol{K}_k} = \boldsymbol{P}_k^ - \boldsymbol{H}_k^{\rm T}{({\boldsymbol{H}_k}\boldsymbol{P}_k^ - \boldsymbol{H}_k^{\rm T} + {\boldsymbol{R}_k})^{ - 1}}
        \)\\
        Step 3.2 Posterior state update\\
        \(
        \hat{\boldsymbol{x}}_{k}^{+}=\hat{\boldsymbol{x}}_{k}^{-} + \boldsymbol{K}_k({y_k} - {h_k}(\hat{\boldsymbol{x}}_{k}^{-}) - {D_k}{u_k})
        \)\\
        Step 3.3 Posterior error covariance update\\
        \(
        \boldsymbol{P}_k^ + = (\boldsymbol{I} - \boldsymbol{K}_k\boldsymbol{H}_k)\boldsymbol{P}_k^ -
        \)
    }
\end{algorithm}

\subsection{Innovation CCM and ACM analysis under OSC error}\label{subsection B}

In \textbf{Algorithm~2}, the expression \({y_k} - {h_k}(\hat{\boldsymbol{x}}_{k}^{-}) - {D_k}{u_k}\) is defined as the Kalman filter innovation. The innovation reflects the difference between the measured terminal voltage and the expected one. Therefore, with a properly designed filter, the innovation should ideally be zero-mean white noise. As the estimation time increases, its covariance should gradually decrease. If the filter innovation is found to be colored, non-zero mean, or has a divergent covariance, it indicates that the measurement equation not only fails to eliminate the cumulative error from the state equation but also causes the estimation result to deviate further from the true value.

In this paper, we consider ${U_{oc,k}} = {f_k}(SO{C_k})$ to be the actual OCV-SOC curve function under the current operating conditions ($T = {T_k}$, ${C_n}$), \(\boldsymbol{H}_k= {\left. {\frac{{\partial {h_k}}}{{\partial \boldsymbol{x}}}} \right|_{\hat{\boldsymbol{x}}_{k}^{-}}} = [{\left. {\frac{{\partial {f_k}(SO{C_k})}}{{\partial SOC}}} \right|_{\hat{\boldsymbol{x}}_{k}^{-} }}, - 1]\) to be the accurate measurement equation partial differential matrix. And ${\widetilde U_{oc,k}} = \widetilde {{f_k}}(SO{C_k})$ represents the original OSC, which was measured either during the battery's historical operation or at room temperature when the battery was newly produced. It is evident that the original OSC has a certain degree of error compared to the actual one. We consider \(\widetilde{\boldsymbol{H}}_{k}= {\left. {\frac{{\partial {{\widetilde h}_k}}}{{\partial \boldsymbol{x}}}} \right|_{\hat{\boldsymbol{x}}_{k}^{-}}} = [{\left. {\frac{{\partial {{\widetilde f}_k}(SO{C_k})}}{{\partial SOC}}} \right|_{\hat{\boldsymbol{x}}_{k}^{-}}}, - 1]\) to be the partial differential matrix of the measurement equation currently used, therefore, ${g}(\boldsymbol{x}_{k}) = {h_k}(\boldsymbol{x}_{k}) - {\widetilde h_k}(\boldsymbol{x}_{k})$ is the gap between the original OSC and the actual OSC, also the difference between the current filter measurement equation and the actual equation.
\begin{equation}\label{Equation 9}
    \begin{split}
    {\widetilde r_k} &= {y_k} - {\widetilde h_k}(\bm{\hat{x}}_{k}^{-}) - {D_k}{u_k} = {h_k}(\boldsymbol{x}_{k}) + {v_k} - {\widetilde h_k}(\bm{\hat{x}}_{k}^{-})\\
& = {\widetilde h_k}(\bm{x}_{k}) - {\widetilde h_k}(\bm{\hat{x}}_{k}^{-}) + {v_k} + {g}(\bm{x}_{k})\\
&= {\bm{\widetilde{H}}_{k}}{\bm{\varepsilon}_k} + {v_k} + {g}(\boldsymbol{x}_{k})
    \end{split}
\end{equation}
where $\bm{\varepsilon}_k=\bm{x}_{k} -\bm{\hat{x}}_{k}^{-}$ is the prior estimation error of step k. When the OSC contains error that varies across different SOC intervals, the mean of the innovation is no longer zero, and it becomes an unsteady time series. Consequently, the following analysis begins with the cross-correlation matrix of the innovation, expressed as \eqref{Equation 10}, assuming $k > i$ for simplicity.
\begin{equation}\label{Equation 10}
    \begin{split}
\mathbb{E}({\widetilde r_k}\widetilde r_i^{\rm T}) &= {\bm{\widetilde{H}}_{k}}(\mathbb{E}({\bm{\varepsilon}_k}\bm{\varepsilon}_i^{\rm T}){\bm{\widetilde{H}}_{i}}^{\rm T} + \mathbb{E}({\bm{\varepsilon}_k}v_i^{\rm T}) + \mathbb{E}({\bm{\varepsilon}_k}){g}{(\bm{x}_{i})^{\rm T}})\\
& + \mathbb{E}({v_k}{(\bm{\widetilde{H}}_{i}{\bm{\varepsilon}_i})^{\rm T}}) + \mathbb{E}({v_k}v_i^{\rm T}) + \mathbb{E}({v_k}){g_i}{(\bm{x}_{i})^{\rm T}}\\
&+ {g}(\bm{x}_{k})\mathbb{E}({\bm{\varepsilon}_i}^{\rm T})\bm{\widetilde{H}}_{i}^{\rm T} + {g}(\bm{x}_{k})\mathbb{E}(v_i^{\rm T}) + {g}(\bm{x}_{k}){g}{(\bm{x}_{i})^{\rm T}}
    \end{split}
\end{equation}
where $v_k$ and $v_i$ are both white noise with a mean of zero, and the past estimation error is independent of the subsequent measurement noise, therefore $v_k$ and $\bm{\varepsilon}_i$ are independent when $k > i$. Therefore, \eqref{Equation 10} can be simplified to
\begin{equation}\label{Equation 11}
    \begin{split}
\mathbb{E}({\widetilde r_k}\widetilde r_i^{\rm T}) &= {\bm{\widetilde{H}}_{k}}(\mathbb{E}({\bm{\varepsilon}_k}\bm{\varepsilon}_i^{\rm T})\bm{\widetilde{H}}_{i}^{\rm T} + \mathbb{E}({\bm{\varepsilon}_k}v_i^{\rm T})) + {\bm{\widetilde{H}}_{k}}\mathbb{E}({\bm{\varepsilon}_k}){g}{(\bm{x}_{i})^{\rm T}}\\
& + {g}(\bm{x}_{k})\mathbb{E}(\bm{\varepsilon}_i^{\rm T})\bm{\widetilde{H}}_{i}^{\rm T} + {g}(\bm{x}_{k}){g}{(\bm{x}_{i})^{\rm T}}
    \end{split}
\end{equation}
 
To further simplify the above formula by calculating $\mathbb{E}({\bm{\varepsilon}_k}\bm{\varepsilon}_i^{\rm T})$, we first explore how the prior estimation error changes under the condition that the Kalman Filter measurement model has error. From \textbf{Algorithm~2}, the one-step state change equation of the EKF can be expressed as
\begin{equation}\label{Equation 12}
\bm{\hat{x}}_{k+1}^{-}=\bm{F}_{k}\bm{\hat{x}}_{k}^{-}+\bm{F}_{k}\bm{\widetilde K}_k\widetilde r_k+\bm{G}_{k}u_k
\end{equation}

Substituting \eqref{Equation 9} into the above equation, then we can get the expression of the $(k+1)$th step's prior estimation error:
\begin{equation}\label{Equation 13}
{\bm{\varepsilon}_{k+1}} = {\bm{\widetilde\psi}_k}{\bm{\varepsilon}_k} + {\bm{\widetilde\xi}_k}
\end{equation}

\noindent where ${\bm{\widetilde\psi}_k} = {\bm{F}_k}(\bm{I} - {\bm{\widetilde K}_k}{\bm{\widetilde H}_k})$, ${\bm{\widetilde\xi}_k} = {\bm{w}_k} - {\bm{F}_k}{\bm{\widetilde{K}}_{k}}({v_k} + {g}(\bm{x}_{k}))$, the above equation can be understood as a linear discrete state equation about $\bm{\varepsilon}_k$. From \eqref{Equation 13} we can conclude that, if $\mathbb{E}(\bm{\varepsilon}_k) = 0$, the prior estimate at $k$th step is an unbiased estimate. Due to the measurement model error ${g}(\bm{x}_{k})\ne 0$, prior estimation error at $k$th step $\mathbb{E}(\bm{\varepsilon}_{k+1}) \ne 0$, that is the prior estimate becomes biased, therefore EKF estimation result is not optimal. From \eqref{Equation 13}, we can deduce the cross-correlation matrix between $\bm{\varepsilon}_k$ and $\bm{\varepsilon}_i$ :
\begin{equation}\label{Equation 14}
\mathbb{E}({\bm\varepsilon _k}\bm\varepsilon _i^{\rm T}) = \mathbb{E}[({\bm{\widetilde\psi} _{k,i}}{\bm{\varepsilon} _i} + \sum\limits_{j = i}^{k - 1} {{\bm{\widetilde \psi}_{k,j + 1}}} {\bm{\widetilde \xi}_j})\bm{\varepsilon}_i^{\rm T}]
\end{equation}

\noindent where $\bm{\widetilde\psi}_{k,i}=\prod\limits_i^{k - 1} {{{\bm{\widetilde \psi} }_j}} $ when $k>i$, and we can also derive $\bm{\widetilde\psi}_{k,i}=\bm I$ when $k=i$. The process error, measurement error of $\bm{\widetilde \xi}_j$ all occur at $i$th step and later, so their cross-correlation matrix with $\bm{\varepsilon}_i^{\rm T}$ equal zero. Therefore, \eqref{Equation 14} can be transformed into
\begin{equation}\label{Equation 15}
\mathbb{E}({\bm\varepsilon _k}\bm\varepsilon _i^{\rm T}) =\bm{\widetilde\psi}_{k,i}\mathbb{E}({\bm\varepsilon _i}\bm\varepsilon _i^{\rm T})- \sum\limits_{j = i}^{k - 1} ({{\bm{\widetilde \psi}_{k,j + 1}}}\bm{F}_j\bm{\widetilde{K}}_{j}g(\bm{x}_{j}))\mathbb{E}(\bm\varepsilon _i^{\rm T})
\end{equation}

In the expression \eqref{Equation 13} of $\bm\varepsilon _k$, only the noise $v_i$ in $\bm{\widetilde \xi}_i$ is related to the $v_i$, then we can get:
\begin{equation}\label{Equation 16}
\mathbb{E}({\bm\varepsilon _k}v_i^{\rm T}) =-\bm{\widetilde\psi}_{k,i+1}\bm{F}_{i}\bm{\widetilde{K}}_{i}R_i
\end{equation}

Substituting \eqref{Equation 15} and \eqref{Equation 16} into the first term on the right side of the \eqref{Equation 11}, we can obtain 
\begin{equation}\label{Equation 17}
\begin{split}
{\bm{\widetilde{H}}_{k}}(\mathbb{E}({\bm{\varepsilon}_k}\bm{\varepsilon}_i^{\rm T})\bm{\widetilde{H}}_{i}^{\rm T} + \mathbb{E}({\bm{\varepsilon}_k}v_i^{\rm T}))={\bm{\widetilde{H}}_{k}}[\bm{\widetilde\psi}_{k,i}\mathbb{E}({\bm\varepsilon _i}\bm\varepsilon _i^{\rm T})\bm{\widetilde{H}}_{i}^{\rm T} \\- \sum\limits_{j = i}^{k - 1}({{\bm{\widetilde \psi}_{k,j + 1}}}\bm{F}_j\bm{\widetilde{K}}_{j}g(\bm{x}_{j}))\mathbb{E}(\bm\varepsilon _i^{\rm T})\bm{\widetilde{H}}_{i}^{\rm T}-\bm{\widetilde\psi}_{k,i+1}\bm{F}_{i}\bm{\widetilde{K}}_{i}R_i]
\end{split}
\end{equation}

In the case of biased filter estimation, $\mathbb{E}({\bm\varepsilon _i}\bm\varepsilon _i^{\rm T})=\bm{P}_{i}^{-}+\mathbb{E}({\bm\varepsilon _i})\mathbb{E}({\bm\varepsilon _i}^{\rm T})$, and replace $\bm{\widetilde\psi}_{i}$ with ${\bm{F}_i}(\bm{I} - {\bm{\widetilde K}_i}{\bm{\widetilde H}_i})$, we can get: 
\begin{equation}\label{Equation 18}
\begin{split}
&\bm{\widetilde\psi}_{k,i}\bm{P}_{i}^{-}\bm{\widetilde{H}}_{i}^{\rm T}-\bm{\widetilde\psi}_{k,i+1}\bm{F}_{i}\bm{\widetilde{K}}_{i}R_i=
\bm{\widetilde\psi}_{k,i+1}(\bm{\widetilde\psi}_{i}\bm{P}_{i}^{-}\bm{\widetilde{H}}_{i}^{\rm T}- \bm{F}_{i}\bm{\widetilde{K}}_{i}R_i)\\&={\bm{\widetilde\psi}_{k,i+1}}(\bm{F}_i\bm{P}_i^- \bm{\widetilde{H}}_{i}^{\rm T} - {\bm{F}_i}{\bm{\widetilde{K}}_{i}}(\bm{\widetilde{H}}_{i}\bm{P}_i^- \bm{\widetilde{H}}_{i}^{\rm T} + {R_i}))=0
\end{split}
\end{equation}

From the above formula, we can know that, part of $\mathbb{E}({\bm\varepsilon _i}\bm\varepsilon _i^{\rm T})$ can be offset by $\mathbb{E}({\bm{\varepsilon}_k}v_i^{\rm T})$. From \eqref{Equation 9}, we can conclude $g(\bm{x}_{k})\approx-\bm{\widetilde{H}}_{k}\bm\varepsilon_k$ when $v_k\ll g(\bm{x}_{k})$ and the filter has converged, that is $r_k\ll g(\bm{x}_{k})$. Therefore, (11) can be rewritten as
\begin{equation}\label{Equation 19}
    \begin{split}
\mathbb{E}({\widetilde r_k}\widetilde r_i^{\rm T}) \approx 
&{\bm{\widetilde{H}}_{k}}[\bm{\widetilde\psi}_{k,i}\mathbb{E}({\bm\varepsilon _i})\mathbb{E}({\bm\varepsilon _i}^{\rm T})-\sum\limits_{j = i}^{k - 1} ({{\bm{\widetilde \psi}_{k,j + 1}}}\bm{F}_j\bm{\widetilde{K}}_{j}g(\bm{x}_{j}))\\ &\mathbb{E}(\bm\varepsilon _i^{\rm T})]{\bm{\widetilde{H}}_{i}}^{\rm T}-g(\bm{x}_{i})g(\bm{x}_{k})
    \end{split}
\end{equation}

$\bm{\widetilde\psi}_{k,i}$ decreases as $k-i$ increases, when $k-i \gg \widetilde{K}_{i,1}$, $\bm{\widetilde\psi}_{k,i} \to 0$, $\mathbb{E}({\widetilde r_k}\widetilde r_i^{\rm T})$ is basically determined by $g(\bm{x}_j), j \in(i,k) $. Since ${\bm{\widetilde{H}}_{k}}\sum\limits_{j = i}^{k - 1} \left|({{\bm{\widetilde \psi}_{k,j + 1}}}\bm{F}_j\bm{\widetilde{K}}_{j}g(\bm{x}_{j}))\mathbb{E}(\bm\varepsilon _i^{\rm T})\right|{\bm{\widetilde{H}}_{i}}^{\rm T}>g(\bm{x}_{i})g(\bm{x}_{k})$ when $g(\bm{x}_j)>0, \forall \bm{x}_j $ or $g(\bm{x}_j)<0, \forall \bm{x}_j $, we can conclude that $\mathbb{E}({\widetilde r_k}\widetilde r_i^{\rm T})<0:g(\bm{x}_j)>0, \forall \bm{x}_j \in (i,k)$; $\mathbb{E}({\widetilde r_k}\widetilde r_i^{\rm T})>0:g(\bm{x}_j)<0, \forall \bm{x}_j \in (i,k)$. However, we have overlooked the fact that the value of $g(\bm{x}_j)$ may contain both positive and negative values within the sampling step $i$ to step $k$, but this situation is relatively rare because in the complete charge/discharge process, there are not many points where the used OSC and the real OSC intersect. Therefore, when $k-i \gg \widetilde{K}_{i,1}$, we can almost assume that, $g(\bm{x}_k)>0:\mathbb{E}({\widetilde r_k}\widetilde r_i^{\rm T})<0;g(\bm{x}_k)<0:\mathbb{E}({\widetilde r_k}\widetilde r_i^{\rm T})>0$

Similarly, assume $k=i$, from \eqref{Equation 10}, the ACM of $\widetilde r_k$ can be derived as
\begin{equation}\label{Equation 20}
    \begin{split}
\mathbb{E}({\widetilde r_k}\widetilde r_k^{\rm T})=&\bm{\widetilde{H}}_{k}\bm{P}_k^-\bm{\widetilde{H}}_{k}^{\rm T}+\bm{\widetilde{H}}_{k}\mathbb{E}({\bm\varepsilon _k})\mathbb{E}({\bm\varepsilon _k}^{\rm T})\bm{\widetilde{H}}_{k}^{\rm T}\\+&2g(\bm{x}_k)\bm{\widetilde{H}}_{k}\mathbb{E}({\bm\varepsilon _k})+R_k
    \end{split}
\end{equation}

From the analysis of the CCM and ACM of the EKF innovation, we can conclude that when the OSC error \( g(\bm{x}) \) is zero, the prior and posterior estimates of the filter should be unbiased. The CCM of the innovation should exhibit white noise with zero mean, and its ACM should have a strict quantitative relationship with the prior estimate covariance and the measurement noise, given by \(\bm{\widetilde{H}}_{k}\bm{P}_k^-\bm{\widetilde{H}}_{k}^{\rm T}+R_k\). As the only observable and testable quantity in the EKF, innovation should be used as a tool to assess filter performance and refine the model. In this paper, the innovation, along with its CCM and ACM, will be utilized as a means to evaluate filter performance and enhance the model.

\subsection{AMMKF-Based SOC Estimation Method}
Through the analysis of the CCM and ACM of the innovation, we can determine the direction of deviation between the original OSC and the actual OSC, though the exact magnitude of this deviation remains unknown. Our goal is to obtain an optimal measurement model parameter that can rapidly correct SOC deviations. If the current SOC estimation is accurate, this optimal parameter should correspond to the slope of the true OSC curve. However, if the SOC estimation is inaccurate, the optimal parameter will differ from the slope of the true OSC curve; otherwise, the SOC estimation error would persist.
To achieve this, a group of Kalman filters can be run in parallel over a time interval containing $L$ sampling points, with a total duration of $L\times \Delta t$. Each filter is configured with a different $\bm{H}_k$. The optimal filter is then selected based on a specified filter performance evaluation method, and its estimation result $\bm{\hat{x}}_{0}^{+}$, $\bm{P}_{0}^{+}$ is chosen as the initial state for the next time interval.

In any time interval \(t \in \text{int}_m\), set up \(n\) Kalman filters, each with its measurement equation parameter \(\bm{H}_k\) set to one of \(\{\bm{H}_m^1, \bm{H}_m^2, \dots, \bm{H}_m^n\}\). The value of \(\bm{H}_m^i\) depends on the model error analysis results derived from the innovation CCM and ACM. Given the short duration of the time interval, the slope of the OSC remains largely unchanged, so \(\bm{H}_m^i\) can be considered constant over the interval. According to \eqref{Equation 9}, the estimated measurement \(\hat{y}_k\) obtained using the non-optimal filter parameter \(\bm{H}_m^i\) may, due to the uncertainty of the measurement noise \(v_k\), be closer to the actual measurement \(y_k\) than the estimation obtained using the optimal parameter \(\bm{H}_m^{op}\). However, as the number of sampling points increases, the estimated measurements from the optimal measurement model generally align more closely with the actual measurements. Therefore, it is essential to explore the change in the probability \(\Pr(\bm{H}_m^j | y_k)\) for each filter's measurement equation parameter given a measurement value \(y_k\).

\begin{figure*}[htbp] 
    \setlength{\abovecaptionskip}{-0.1cm}  
    \setlength{\belowcaptionskip}{-0.1cm}   
    \centering
    \includegraphics[width=\textwidth]{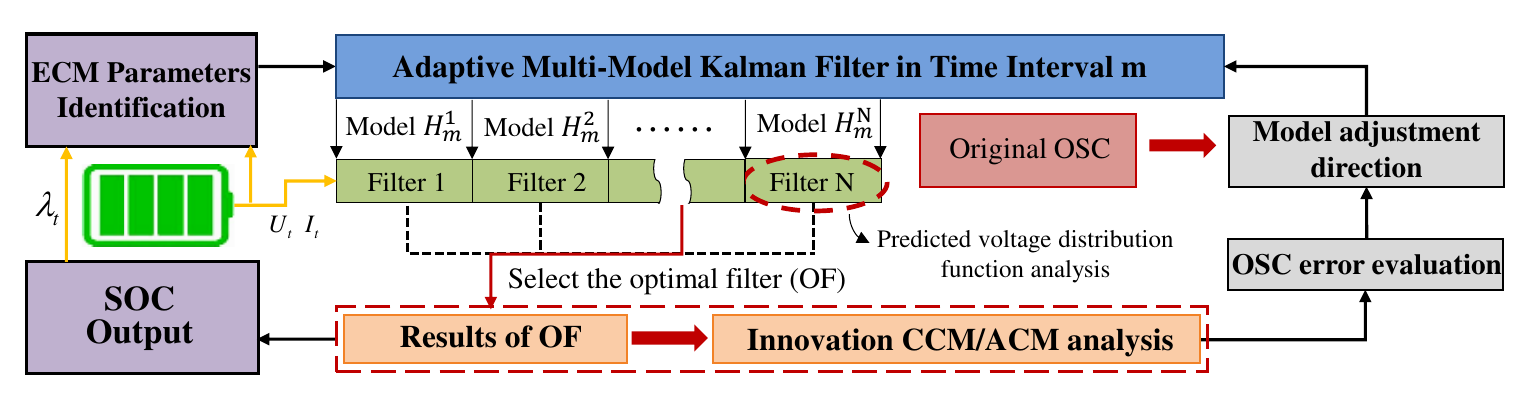}
    \caption{Flowchart of SOC estimation method.}
    \label{Whole}
    \vspace{-0.5cm} 
\end{figure*}

\begin{equation}\label{Equation 21}
\Pr(\bm{H}_m^j\left| {{y_k}} \right.) = \frac{{\Pr ({{\hat y}_k^j = {y_k}\left| \bm{H}_m^j \right.)}\Pr(\bm{H}_m^j)}}{{\sum\limits_{i = 1}^n {\Pr ({{\hat y}_k^j} = {y_k}\left| \bm{H}_m^i \right.)\Pr (\bm{H}_m^i)} }}
\end{equation}
where $\Pr (\bm{H}_m^j)$ is the  probability value of $\bm{H}_m^j$ before the measurement data $y_k$ is passed in, considering all the previous measurement data before $k$th step, therefore $\Pr (\bm{H}_m^j) = \Pr (\bm{H}_m^j\left| {{y_{k - 1}}} \right.)$. ${\hat y}_k^i$ is the estimated measurement value of $i$th filter under $\bm{H}_m^i$ parameter conditions. Since the measurement data $y_k$ is continuous, the probability that any filter's estimated measurement value equals $y_k$ is 0, that is $\Pr ({\hat y_{ki}} = {y_k}\left| {\bm{H}_m^i} \right.)=0$. Therefore, we use the fact that the probability of an event is proportional to its probability density function to rewrite the above formula as
\begin{equation}\label{Equation 22}
\Pr(\bm{H}_m^j\left| {y_k} \right.)=\frac{\mathrm{pdf}({{\hat y}_k^j = {y_k}\left| \bm{H}_m^j \right.)\Pr (\bm{H}_m^j\left|y_{k-1} \right.)}}{\sum\limits_{i = 1}^n {\mathrm{pdf}({{\hat y}_k^i} = {y_k}\left| \bm{H}_m^i \right.)\Pr (\bm{H}_m^i\left|y_{k-1} \right.)} }
\end{equation}

If $\bm{H}_m^j=\bm{H}_m^{op}$ holds true, it means that the filter has most accurate estimation result, that is, $\hat{\bm{x}}_k^{j-}\to\bm{x}_k$. Since the measured value at $k$th step $y_k$ depends on $\bm{x}_k$ and $v_k$, we can get $\mathrm{pdf}({{\hat y}_k^j} = {y_k}\mid\bm{H}_m^j)=\mathrm{pdf}({{\hat y}_k^j} = {y_k}\mid\hat{\bm{x}}_k^{j-})$. And when $\bm{H}_m^j=\bm{H}_m^{op}$, the prior estimate of filter $j$ is unbiased, that is, $\mathbb{E}(\hat{\bm{x}}_k^{j-})=\bm{x}_k\approx \hat{\bm{x}}_k^{j-} $. Taylor expansion of ${\hat y}_k^j$ at the starting point $\bm{\hat{x}}_{0}^{+}$ in time interval $m$, retaining the first differential term, we can obtain the following formula
\begin{equation}\label{Equation 23}
{\hat y}_k^j= \bm{H}_m^j\hat{\bm x}_{k}^{j-} + (h_k^j(\hat{\bm x}_0^ + ) - \bm{H}_m^j\hat{\bm x}_0^ + ) + {D_k}{u_k} + {v_k}
\end{equation}
where $h_k^j$ is the OSC function of filter $j$ at $k \in int_m$. In all the KFs we designed at any time interval, $\bm{w}_k$ and $v_k$ are Gaussian distributed, so if $\bm{H}_m^j=\bm{H}_m^{op}$ holds true, the resulting ${\hat y}_k^j$ is a linear combination of Gaussian random variable, its distribution is also Gaussian. Then $\mathbb{E}({\hat y}_k^j)$ is approximately equal to $\bm{H}_m^j\hat{\bm x}_{k}^{j-} + (h_k^j(\hat{\bm x}_0^ + ) - \bm{H}_m^j\hat{\bm x}_0^ + ) + {D_k}{u_k}$, its variance ${S_k^j} = \bm{H}_m^j\bm{P}_{k}^{j-} {\bm{H}_m^j}^{\rm T} + {R_k}$. In the probability density function of ${\hat y}_k^j$, the probability density of ${\hat y}_k^j=y_k$ can be expressed as
\begin{equation}\label{Equation 2}
\mathrm{pdf}({\hat{y}_k^j} = {y_k}\left| {\bm{H}_m^j} \right.) = \frac{1}{{\sqrt {2\pi S_k^j } }}\exp ( - \frac{{{{({y_k} - \mathbb{E}({{\hat y}_k^j}))}^2}}}{{2{S_k^j}}})
\end{equation}

Based on the above formula, update the probability that each filter's measurement model is optimal after each sample step, denoted as $\Pr(\bm{H}_m^i\left| {{y_k}} \right.)$, $\bm{H}_m^i\in\{\bm{H}_m^1,\bm{H}_m^2,..., \bm{H}_m^n\}$. At the end of $int_m$, the optimal measurement model parameter of $int_m$ is selected based on
$\bm{H}_m^{op}=\bm{\mathop{\arg\min}}_{\bm{H}_m^i}\Pr (\bm{H}_m^i \mid y_k)$. The filter associated with the optimal measurement model parameter will be considered the optimal filter of $int_m$, and it's SOC estimation result of each step is regarded as the final estimation result of the AMMKF during $int_m$. For the subsequent interval \(int_{m+1}\), the initial estimates and error covariances of each filter are set according to the state estimation results and error covariance of the optimal filter at the last sampling point in \(int_m\).

\subsection{AMMKF Parameters Adaptive Adjustment Strategy}
In real filter operation scenarios, typically only one sampling data point is obtained at each sampling time. Since \(\bm{H}_{m-1}^{\text{op}}\) and \(\bm{H}_m^{\text{op}}\) remain constant within \(int_{m-1}\) and \(int_m\), respectively, the dependency structure in the innovation does not change over time. Consequently, the only factor affecting the innovation CCM is the distance between \(\widetilde{r}_{m-1,\text{end}}\) and \(\widetilde{r}_{m,\text{end}}\) in the time series. Therefore, the CCM between the new information at the last sampling point of two adjacent time intervals can be calculated by
\begin{equation}\label{Equation 24}
\mathbb{E}({\widetilde r_{m - 1,end}} \widetilde r_{m,end}^{\rm T}) = \frac{1}{L}\sum\limits_{i = 1}^L {{{\widetilde r}_{m - 1,i}}} \widetilde r_{m,i}^{\rm T}
\end{equation}

From the analysis in \ref{subsection B}, we can determine that if $\mathbb{E}({\widetilde r_{m - 1,end}} \widetilde r_{m,end}^{\rm T})>0$, then $g(\bm{x}_{m,end})<0$. Conversely, if \(\mathbb{E}(\widetilde{r}_{m-1,\text{end}} \widetilde{r}_{m,\text{end}}^{\rm T}) < 0\), then \(g(\bm{x}_{m,\text{end}}) > 0\). In the time interval \(int_{m+1}\), the filter parameters need to be adjusted based on the sign of \(g(\bm{x}_{m,\text{end}})\). For the discharge process, if \(g(\bm{x}_{m,\text{end}}) < 0\), it indicates that the OSC we are using is higher than the actual OSC at sampling point \(\bm{x}_{m,\text{end}}\). From \eqref{Equation 9}, we can also infer that the estimated SOC of the optimal filter in \(int_m\) is lower than the real SOC. Therefore, in the next time interval \(int_{m+1}\), we should set the AMMKF parameters \(\{\bm{H}_{m+1}^1, \bm{H}_{m+1}^2, \ldots, \bm{H}_{m+1}^n\}\) to be larger than \(\bm{\widetilde{H}}_{m+1}\) to achieve a rapid decrease in OSC and a slower decrease in the estimated SOC. Conversely, when \(g(\bm{x}_{m,\text{end}}) > 0\), set \(\{\bm{H}_{m+1}^1, \bm{H}_{m+1}^2, \ldots, \bm{H}_{m+1}^n\}\) to be smaller than \(\bm{\widetilde{H}}_{m+1}\) to achieve a rapid decrease in SOC and a slower decrease in OSC. The parameter adjustment strategy for the charging process is the reverse of the above.

\section{Results and discussions}\label{Results}
In this section, the experimental data of $\mathrm{LiFePO_4}$ battery from the Center for Advanced Life Cycle Engineering (CALCE)~\cite{web} battery group is used to verify and analyze the proposed SOC estimation method, ~\cite{2014AE} describes the experiments conducted to generating the data. Table~\ref{Table 1} presents the key specifications of the battery. The data set includes low-current charge and discharge data and dynamic configuration data of the battery at various temperatures (-10℃, 0℃, 10℃, 20℃, 25℃, 30℃, 40℃, 50℃). The low-current charge and discharge experiment uses a current of 0.05A (about 1/20C) with a sampling interval of 5s. The dynamic configuration file includes three battery cycling tests: Dynamic Stress Test(DST), Federal Urban Driving Schedule (FUDS), Supplemental Federal Test Procedure (US06).

\begin{table}[htbp]
    \vspace{-0.3cm} 
    \setlength{\abovecaptionskip}{-0.1cm}
    \setlength{\belowcaptionskip}{-0.1cm}
    \renewcommand\arraystretch{1.5} 
    \setlength{\tabcolsep}{0.12cm} 
    \caption{Key specifications of the battery used for proposed SOC estimation method performance test}\label{Table 1}
    \begin{center}
    \begin{tabular}{cccccccc}
    \toprule 
    & \shortstack{Cathode\\Materials} 
    & \shortstack{Rated\\Capacity} 
    & \shortstack{Actual\\Capacity} 
    & \shortstack{Charge\\Cut-off Voltage} 
    & \shortstack{Discharge\\Cut-off Voltage} \\ 
    \midrule
    & $\mathrm{LiFePO_4}$ & 1.1Ah & 1.063Ah & 3.6V & 2.0V \\
    \bottomrule   
    \end{tabular}
    \end{center}
    \vspace{-0.7cm} 
\end{table}	

\subsection{ECM Parameters Identification Results and OSC Characteristics Analysis}
\begin{figure}[htbp] 
    \setlength{\abovecaptionskip}{-0.1cm}  
    \setlength{\belowcaptionskip}{-0.1cm}   
    \centerline{\includegraphics[width=0.95\columnwidth]{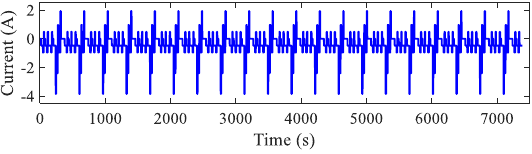}}
    \caption{Current profile of entire DST test}
    \label{DST_Current}
    \vspace{-0.6cm}
\end{figure}

\begin{figure}[htbp] 
    \setlength{\abovecaptionskip}{-0.1cm}  
    \setlength{\belowcaptionskip}{-0.1cm}   
    \centerline{\includegraphics[width=1\columnwidth]{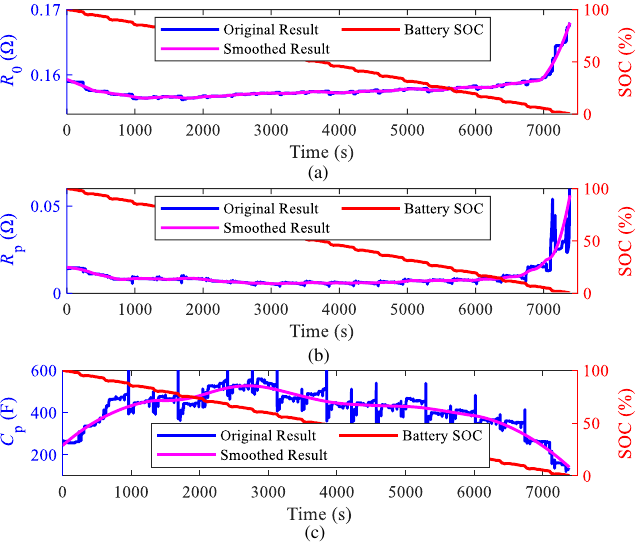}}
    \caption{ECM parameters identification results: (a) $R_0$, (b) $R_p$, (c) $C_p$}
    \label{R0RPCP}
    \vspace{-0.4cm} 
\end{figure}

The ARLS algorithm presented in this paper is employed to estimate the ECM parameters of the battery at different SOC levels under DST. Fig.~\ref{DST_Current} illustrates the excitation current data of DST, which consists of twenty sets of repeated charge and discharge excitations, lasting approximately two hours, ultimately causing the battery SOC to drop from 100\% to 0. The DST sampling interval is 1s. Fig.~\ref{R0RPCP} presents the numerical variations of \(R_0\), \(R_p\), and \(C_p\) in the ECM of the tested battery under the operating condition of 25°C. As seen in the figure, as the battery SOC decreases, the ohmic internal resistance \(R_0\) and the polarization internal resistance \(R_p\) generally exhibit a trend of first decreasing and then increasing. However, \(R_0\) is not significantly affected by SOC, while \(R_p\) increases markedly when the battery is at a very low SOC. As the SOC decreases from 100\%, \(C_p\) initially increases and then decreases.

\vspace{-0.2cm}

\begin{figure}[htbp] 
    \vspace{-0.4cm} 
    \setlength{\abovecaptionskip}{-0.1cm}  
    \setlength{\belowcaptionskip}{-0.1cm}   
    \centerline{\includegraphics[width=0.95\columnwidth]{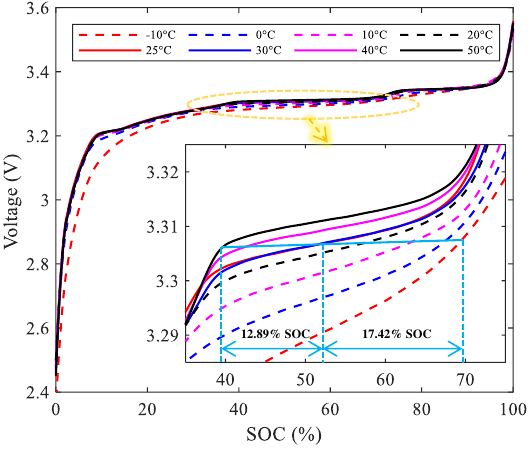}}
    \caption{OSCs of tested battery within [-10℃,50℃]}
    \label{OSCs}
    \vspace{-0.8cm} 
\end{figure}

Fig.~\ref{OSCs} shows the OSCs at various temperatures. The impact of temperature on the OSC of the \(\mathrm{LiFePO_4}\) battery is significant, with the shape of the OSC changing markedly under low-temperature conditions. 
Comparing the OSCs in the second plateau period, the SOC difference between the 50°C condition and the 25°C condition under the same OCV is 12.89\%, and the SOC difference between the 25°C condition and the -10°C condition is 17.42\%. This indicates that if the Kalman filter's measurement equation uses the 25°C OSC, even without any noise in the measurement equipment, the estimated SOC will deviate by approximately +0.51\%/°C as the battery temperature rises due to heat accumulation. Conversely, operation under low-temperature conditions will result in an SOC estimation error of about -0.43\%/°C. During the non-plateau region, the OSC slope is significantly lower at low temperatures than at room temperature. This slope error directly affects the Kalman filter gain, and combined with the differences in curve values, makes it difficult for the filter to converge, leading to large output and covariance values.

\begin{figure*}[t] 
    \setlength{\abovecaptionskip}{-0.1cm}  
    \setlength{\belowcaptionskip}{-0.1cm}   
    \centering
    \includegraphics[width=\textwidth]{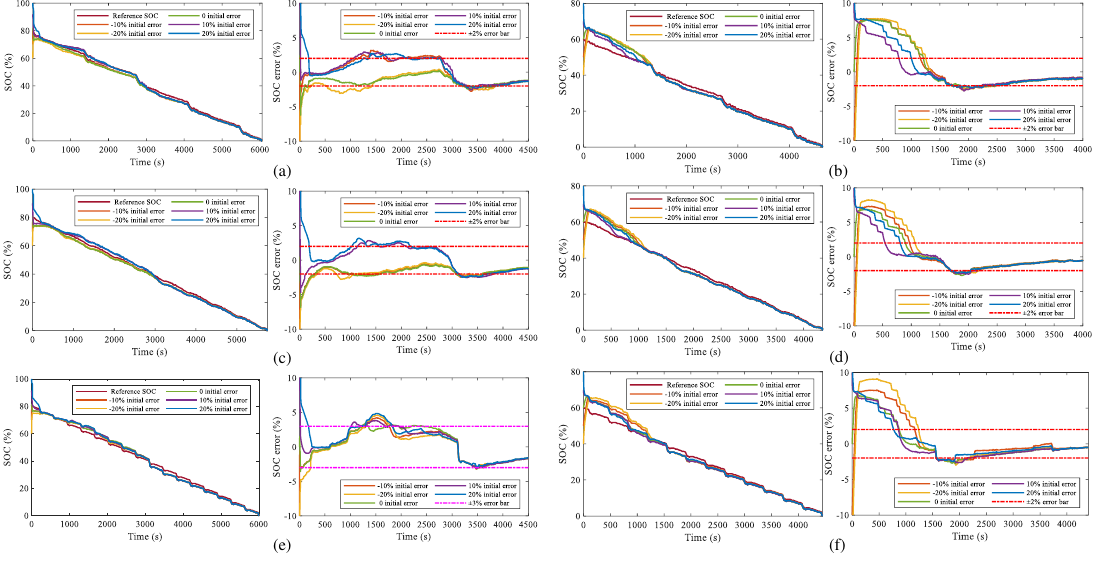}
    \caption{SOC estimation results of $\mathrm{LiFePO_4}$ operating at 50℃ based on 25℃ OSC: (a)FUDS, 80\% initial SOC, (b)FUDS, 60\% initial SOC, (c)US06, 80\% initial SOC, (d)US06, 80\% initial SOC, (e)DST, 80\% initial SOC, (f)DST, 60\% initial SOC}
    \label{SOC_50}
    \vspace{-0.1cm}
\end{figure*}
\begin{figure*}[htbp] 
    \setlength{\abovecaptionskip}{-0.1cm}  
    \setlength{\belowcaptionskip}{-0.1cm}   
    \centering
    \includegraphics[width=\textwidth]{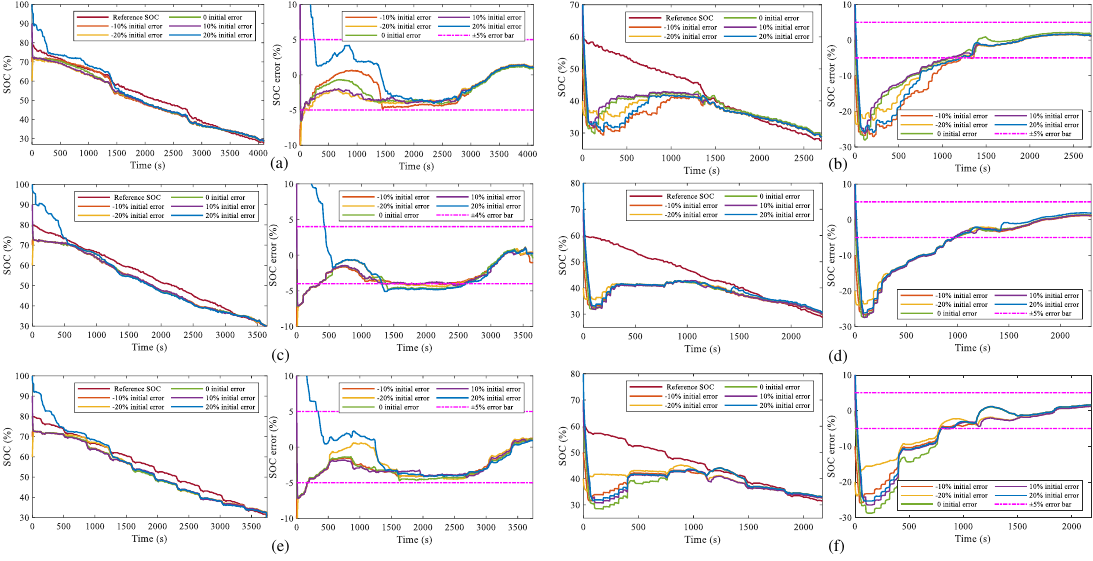}
    \caption{SOC estimation results of $\mathrm{LiFePO_4}$ operating at -10℃ based on 25℃ OSC: (a)FUDS, 80\% initial SOC, (b)FUDS, 60\% initial SOC, (c)US06, 80\% initial SOC,minus (d)US06, 80\% initial SOC, (e)DST, 80\% initial SOC, (f)DST, 60\% initial SOC}
    \label{SOC_-10}
    \vspace{-0.3cm}
\end{figure*}

\subsection{Estimation Accuracy and Robustness Test of the Proposed Method}
To verify the performance and generalization ability of the algorithm under different operating conditions, tests were conducted using FUDS, US06, and DST profiles. In the early stages of estimation, all measurement models in the AMMKF rely entirely on the original OSC (25°C) until the filter innovation converges. Once innovation convergence is achieved, the parameters of each filter in the AMMKF are adjusted based on the CCM and ACM analysis results. We considered two extreme scenarios, where the OSC shifts up and down most within the plateau region. The ECM parameters are derived from the ARLS results based on real-time operating data, ensuring that the state model is accurate.

Fig.~\ref{SOC_50} presents the results of SOC estimation starting from 80\% and 60\% initial SOC with initial SOC errors ranging from -20\% to 20\%, under the 50°C FUDS, US06, and DST. The data after 80\% and 60\% SOC were used for testing the estimation method, so the actual initial polarization voltage is not zero (corresponding to a -100\% initial error). The data sampling interval is 1 second, and 20 seconds is selected as the AMMKF update interval. According to the results in Fig.~\ref{SOC_50}, with an initial SOC of 80\%, the SOC estimation results from the AMMKF can be quickly corrected to within a 5\% error across all three tests and under various initial SOC errors. In the OSC plateau region, the SOC estimation error increases due to the small slope of the OSC. However, after passing through the plateau region, the SOC estimation error decreases and converges to within 2\%. Under the condition of 60\% initial SOC, it takes longer to achieve SOC error convergence because the initial polarization voltage estimation error within the OSC plateau has a significant impact on the SOC estimation results.

Fig.~\ref{SOC_-10} illustrates the SOC estimation results at -10°C for initial SOCs of 80\% and 60\%, under FUDS, US06, and DST tests. Due to the significant increase in ECM parameters at this temperature, the battery terminal voltage reaches the discharge cut-off voltage of 2.0V when there is still approximately 30\% SOC remaining. It can be observed that, compared with the estimation results at 50°C, the results at -10°C take longer to converge and exhibit poorer accuracy in the plateau region. This discrepancy is attributed to the substantial differences in both the values and slopes of the OCV between 25°C and -10°C. The results shown in (b), (d), and (e) indicate that at 60\% SOC, the OCV of the battery at 25°C is higher than the OCV at -10°C. Combined with the initial polarization voltage error, this causes the SOC estimate in the early stages (before OCV correction) to converge to a value significantly lower than the reference, resulting in an SOC error of approximately -30\%. Although the measurement model parameters of each filter in the AMMKF are increased in subsequent time intervals after CCM and ACM analysis, it still requires considerable time for the SOC estimation error to converge within 5\%. Consequently, the proposed method demonstrates a superior downward adjustment effect on SOC compared to its upward adjustment capability. It is noteworthy that when there is a large initial polarization voltage error, the estimation process with a more accurate initial SOC estimate does not necessarily achieve better convergence and accuracy. This is because, during the first estimation interval, the algorithm relies entirely on the original 25°C OCV. To minimize innovation, the filter may cause the SOC to deviate further from the reference value, which in turn significantly influences the OCV trend constructed in each subsequent interval. By combining the results from Fig.~\ref{SOC_50} and Fig.~\ref{SOC_-10}, it can be seen that the algorithm proposed in this paper, based on the 25°C OCV, achieves higher accuracy and faster convergence for battery SOC estimation at extreme operating temperatures of 50°C and -10°C. It also exhibits strong robustness to initial SOC errors.

\subsection{Comparision of results based on proposed method and UKF}\label{Comparision}
\begin{figure}[htbp] 
    \vspace{-0.3cm}
    \setlength{\abovecaptionskip}{-0.1cm}  
    \setlength{\belowcaptionskip}{-0.1cm}   
    \centerline{\includegraphics[width=1\columnwidth]{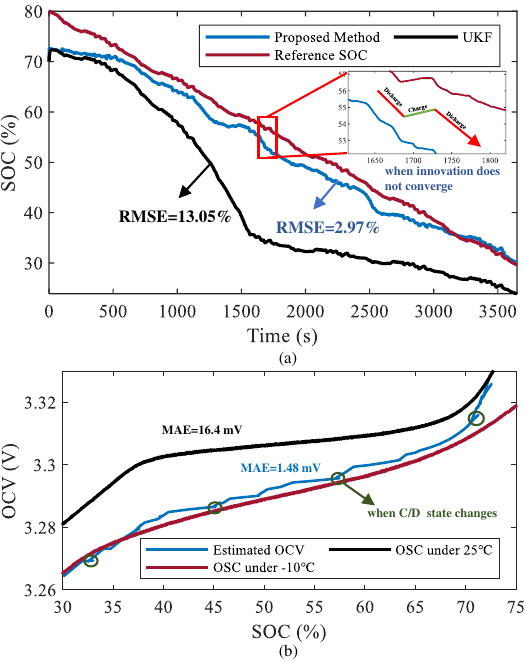}}
    \caption{Comparison of UKF and proposed method estimation result based on 25℃ OSC under -10℃ US06 condition: (a)comparison of SOC result, (b) the OSC corrected by proposed method}
    \label{Compare}
    \vspace{-0.5cm} 
\end{figure}

Fig.~\ref{Compare}(a) presents a comparison of the SOC estimation results between the UKF and the proposed method based on the 25℃ 
OSC, under the -10℃ US06 test. The initial SOC is set to 80\% with a -10\% initial SOC error for both methods. The SOC estimated by the UKF exhibits a significant error throughout the entire experimental range, with a maximum error of -24.72\% and an RMS error of 13.05\%. In contrast, the proposed method, where the time interval length of the AMMKF is set to 50s, shows a much better performance. The SOC estimation result converges quickly to within a 3\% error but then becomes slightly larger in the plateau region, with a maximum error of -4.68\%. The RMS error for the entire experimental process is 2.97\%, which is significantly lower than that of the UKF method. When the innovation does not converge, the correction effect of the posterior estimate on the SOC is much greater than the update from the prior estimate, which can cause scenarios where the battery is being charged but the estimated SOC is decreasing.
Fig.~\ref{Compare}(b) displays the OSC followed by the optimal filters of the AMMKF. It can be observed that in the initial time interval, the estimation relies entirely on the 25℃ OSC. As time progresses, in subsequent intervals, the OSC followed by the optimal filter gradually deviates from the 25℃ OSC and approaches the -10℃ OSC. The mean absolute error (MAE) between the 25℃ OSC and the -10℃ OSC is 16.4 mV, while the average error between the AMMKF-corrected OSC and the -10℃ OSC is just 1.48 mV. Notably, the same SOC in the corrected OSC may correspond to two different OCV values, as highlighted by the green circle in Fig.~\ref{Compare}(b). This occurs because the US06 test condition includes both charging and discharging processes. The measurement models of the optimal filters differ between adjacent time intervals, so when the process transitions from discharging to charging, the OCV value corresponding to the same SOC changes. However, this impact is minimal due to the short time interval length of the AMMKF.

\begin{figure}[htbp] 
    \setlength{\abovecaptionskip}{-0.1cm}  
    \setlength{\belowcaptionskip}{-0.1cm}   
    \centerline{\includegraphics[width=1\columnwidth]{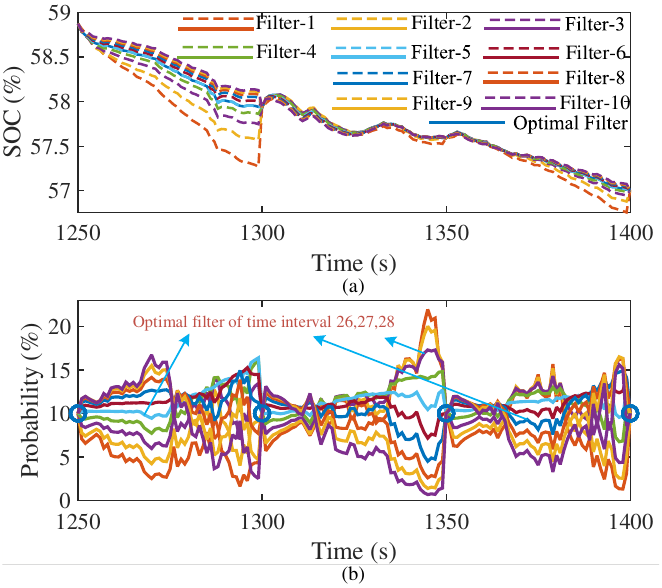}}
    \caption{AMMKF details during time intervals: (a) Output results of each filter in AMMKF, (b) Probability update of each filter in AMMKF}
    \label{Details}
    \vspace{-0.5cm} 
\end{figure}

Fig.~\ref{Details} shows the details of the AMMKF during the time intervals from 1250s to 1400s (interval 26, 27, and 28). Fig.~\ref{Details}(a) presents the SOC estimation results of each filter within these intervals. The differences in the output values of each filter within a time interval are influenced by the current. Fig.~\ref{Details}(b) illustrates the probability update process for each filter. Filters 5, 3, and 7 are selected as the optimal filters for intervals 26, 27, and 28, respectively, due to having the highest probability at the end of these intervals. It can be observed that the filter with the smallest estimation error is not necessarily chosen as the optimal filter. This is because the selection of the optimal filter is based on the probability distribution function of its predicted voltage, choosing the model parameters that best match the real data. During the time interval, the parameter slope of OSC is not updated. Consequently, if the interval is set too long, none of the filters will achieve high precision. Conversely, if the interval is too short, CCM and ACM analyses will be inaccurate, and the probability assignments of the filters will be significantly affected by noise.

\section{Conclusion}\label{Conclusion}
In this paper, a $\mathrm{LiFePO_4}$ battery SOC estimation method based on AMMKF that accepts OSC with error is proposed.
By analyzing the CCM and ACM of KF innovation, it is concluded that when there is an error in OSC, the sign of the CCM of two sampling points far apart is inversely related to the sign of the OSC error, and the ACM of the sampling point will also be higher than the theoretical value. Based on this, the measurement equation parameters of AMMKF are dynamically adjusted upward or downward compared to the original OSC, and the probability of each filter is assigned according to the distribution of predicted voltage values, then the optimal filter is selected.

The performance of the proposed algorithm is tested under the two cases where the existing OSC shifts up and down most in the plateau regon. The results show that the proposed method has high accuracy and robustness performance under three tests, and the RMSE of estimated SOC is less than 3\%, which is more than 10\% less than the traditional method. Subsequent research will focus on the impact of the initial polarization voltage estimation error on the corrected OSC trend and reduce the convergence time of the estimation result.

\bibliographystyle{IEEEtran}
\bibliography{IEEEabrv,SOC_Estimation}

\end{document}